\documentclass{article}%
\usepackage{amsmath}%
\usepackage{amsfonts}%
\usepackage{amssymb}%
\usepackage{graphicx}

\begin{document}

\title{A New Continuum Formulation for Materials--Part I. The Equations of Motion for
a Single-Component Fluid}
\author{Melissa Morris\\3420 Campus Blvd. NE\\Albuquerque, NM 87106\\\ mmorris400@comcast.net}
\date{January 18, 2017}
\maketitle

\begin{abstract}
The continuum equations of fluid mechanics are rederived with the intention of
\ keeping certain mechanical and thermodynamic concepts separate. \ A new
"mechanical" mass density is created to be used in computing inertial
quantities, whereas the actual mass density is treated as a thermodynamic
variable. \ A new set of balance laws is proposed, including a mass balance
equation with a non-convective flux. \ The basic principles of irreversible
thermodynamics are used to obtain linear constitutive equations that are
expansions of--not only the usual affinities involving gradients of
temperature and velocity--but also the gradient of the chemical potential.
\ Transport coefficients are then chosen based on an elementary diffusion
model, which yields simple constitutive laws featuring just two transport
parameters: one for the longitudinal part of the motion and one for the
rotational part. \ The resulting formulation differs from the
Navier-Stokes-Fourier equations of fluid motion. \ In order to highlight key
similarities and differences between the two approaches, several examples in
fluid mechanics are treated in \textsc{part II}, including sound propagation,
light scattering, steady-state shock waves, thermophoresis, and Poiseuille flow.\ 

\end{abstract}
\tableofcontents

\section{\label{secin}Introduction}

The starting point for the theory proposed here is simple but somewhat
abstract. \ Instead of the five equations and five unknowns that the
Navier-Stokes-Fourier (NSF) formulation employs to describe a single-component
fluid in three spatial dimensions, this formulation requires six (although the
number of equations/unknowns drops back down to five for most linearized
problems). \ The extra equation governs a new quantity, which I call the
mechanical mass density and label as $\overline{m} $. \ It is a book-keeping
variable to be used when defining inertial quantities such as linear and
angular momentum and kinetic energy.

The mechanical mass density is assumed to satisfy a continuity equation with
nothing but convection to move it, while the (actual) mass density $m$ is
governed by a balance law that includes a mass diffusion term in addition to a
convection term.

Behind the equations of motion that follow, a philosophy has been adopted\ to
keep thermodynamics and mechanics separate in some key respects. \ The
non-convective mass flux, $\underline{q}_{M}$, is treated as one would the
non-convective heat or internal energy flux, not, in other words, as a
momentum involving a diffusion velocity. \ It is assumed to be dissipative
and, consequently, its form is chosen based on non-equilibrium thermodynamic
principles. \ On the other hand, the mechanical mass density may be viewed as
the background mass density the material would have if it were somehow in a
state of thermodynamic equilibrium with only the velocity field directly
affecting it through the laws of continuum mechanics. \ The balance laws
stemming from these ideas are proposed in \S \ref{secbl}. \ In them, there
appears a local continuum mechanical velocity, $\underline{v}$, which is used
in the convection terms and in the definitions of the momentum and kinetic
energy densities. \ Also, in these inertial density definitions, the
mechanical mass density is employed instead of the actual mass density.
\ Finally, when postulating a form for the total pressure tensor appearing in
the momentum equation, there is no allowance for mass diffusion effects, nor
are mass diffusion and external body forces allowed to affect one another directly.

The above statements warrant clarification:\ I am not claiming there to be an
absence of inertial effects arising from the mass diffusion. \ After all, one
may define momentum and kinetic energy densities involving the actual mass
density or the total mass flux and then postulate balance laws for the time
rate of change of those quantities. \ I prefer not to work with these,
however, as their balance laws are complicated by cross-effects. \ Within my
formulation, the velocity $\underline{v}$\ is not responsible for the overall
mass flow. \ Instead, it is seen to govern, through convection, only the
mechanical part of the flow, e.g. that which is caused by pressure
disturbances, moving boundaries, gravity, etc., whereas the diffusional part
of the mass flow is handled separately in accordance with non-equilibrium
thermodynamics. \ With this in mind, the balance laws given in \S \ref{secbl}%
\ for the linear momentum, angular momentum, and kinetic energy densities are
understood to be laws for the \textit{mechanical part} of the linear momentum,
angular momentum, and kinetic energy densities. \ There are some interesting
benefits to this approach, a few of which are (1)\ it circumvents the angular
momentum conservation issues and other concerns raised in \"{O}ttinger et al.
\cite{ot}--see \S 3, (2) because the diffusional part of the mass flux can be
made to cancel with the convective part, it allows us to model an impermeable
boundary with a non-zero normal velocity at that boundary as arising in the
case of non-infinite acoustic impedance at a sound barrier--see \S 6 of
\textsc{part II}, and (3) after appropriate constitutive equations and
transport coefficients are chosen, it leads to the case of pure diffusion via
Fick's law when mass density gradients are present yet conditions have been
chosen to eliminate all mechanical forces--see \S 4 of \textsc{part II}.

In \S \ref{secce} and \textsc{appendix} \ref{applc} of the current paper, the
principles of irreversible thermodynamics are used to select general forms for
the dissipative fluxes appearing in the balance laws. \ The NSF formulation
possesses two dissipative fluxes, the heat flux and the viscous pressure
tensor, whereas mine at the outset possesses four: the aforementioned two,
plus a non-convective mass flux and an energy flux due to chemical work. \ I
choose expressions for the chemical energy and heat fluxes in strict analogy
to the first law of equilibrium thermodynamics. \ With standard techniques, my
formulation produces linear constitutive laws that are expanded in terms of
the usual affinities involving gradients of the velocity and temperature, as
in the case of the NSF formulation. \ However, there arise additional terms
involving gradients of the chemical potential. \ Since the non-convective
internal energy flux may be expressed as the sum of those due to heat and
chemical work, my formulation is found to have three independent constitutive
laws and six transport coefficients (two for each law), five of which are
independent if Onsager reciprocity is enforced.\footnote{It should be noted
that at this stage, one could choose the coefficients in my formulation such
that the non-convective mass flux vanishes, thereby yielding the NSF
formulation as a special case, with its two constitutive laws and three
transport parameters (bulk and shear viscosity and heat conductivity).} \ It
may be difficult, if not impossible, to devise experiments for measuring each
of my general transport coefficients individually. \ Therefore, the proposed
framework would not be very convenient or useful without a transport model to
guide specific choices for the formulas of these transport coefficients.

Many thermodynamics textbooks, e.g. Reif \cite[\textsc{ch. }12]{reif}, present
a simple particle diffusion model as an introduction to non-equilibrium
transport processes, the general idea being that the particles of a gas
diffuse via random collisions with one another, and they carry with them
extensive quantities, such as momentum and energy, which are exchanged during
the collisions. \ These types of models, constructed for a classical monatomic
ideal gas, are found to be qualitatively useful but quantitatively inaccurate,
and thus they are discarded for more intricate procedures based on the
Boltzmann equation. \ However, within the setting of my general formulation,
it is of interest to revisit these ideas of diffusion. \ I claim that my
construction helps make such an elementary picture a viable way of describing
non-equilibrium transport in fluids. \ Moreover, the ideas behind this model
are not constrained to the case of a classical monatomic ideal gas. \ First,
in \S \ref{seclin} I linearize the equations of motion and decompose the
problem into its longitudinal and rotational parts. \ Next, in \S \ref{sectc},
I suggest a transport model in which each of the quantities appearing in the
longitudinal part, i.e. the mass, internal energy, and longitudinal momentum,
all diffuse with the same coefficient $D$.\footnote{\label{fnsd}Note that
$D$\ is not the self-diffusion coefficient, although these quantities are the
same order of magnitude for gases (see \textsc{appendix C} of \textsc{part II}
).} \ Furthermore, since the linearized versions of NSF and my formulation
yield identical rotational parts, which consist of one equation governing the
rotational momentum that may be uncoupled from the rest of the equations, I
postulate that the shear viscosity parameter $\eta$ is the same for both
descriptions. \ This gives rise to formulas for all six of the transport
coefficients involving just two parameters, and Onsager reciprocity is
automatically satisfied. \ The heat flux law arising from such a model turns
out to be different from Fourier's, and the total viscous pressure resulting
from my assumptions differs from that of Navier-Stokes, as well. \ Of course,
it is important to be able relate $D$ to the usual transport parameters, the
bulk and shear viscosities and heat conductivity, in problems for which the
NSF formulation is known to give the right answer. \ This is easily done by
using sound attenuation data, for example, as shown in \S 5 of \textsc{part
II}. \ An equally important issue is that there must be a way of
experimentally distinguishing between the two theories for a problem where the
NSF equations can be shown to fail. \ Light scattering, addressed in \S 7 of
\textsc{part II},\ may prove to be such an experiment, and it is currently
being conducted.

When considering the dimensionless Knudsen number, $K\!n$, defined as the
ratio of the fluid's mean free path length to some characteristic length scale
in a particular problem, it should be emphasized that, like NSF, mine is a
formulation intended to be used in the hydrodynamic, or small $K\!n$, regime.
\ I do not make any general claims that my continuum formulation works well
into the more rarefied gas regime, although experimental data for sound
propagation in the noble gases shows that mine performs significantly better
than Navier-Stokes-Fourier there--see \textsc{figure 1} of \textsc{part II}.
\ The previously mentioned light scattering experiment is being designed to
study gases fully in the hydrodynamic regime which, up until now, has either
been ignored or not adequately resolved in all other such experiments.

In addition to sound propagation and light scattering, several other
applications are explored in \textsc{part II}. \ As a general rule, one finds
that the formulas derived from my equations tend to be, in many ways, much
simpler and easier to interpret than those obtained from the NSF theory. \ In
fact for some examples, when only approximate answers may be computed with the
Navier-Stokes-Fourier equations, exact solutions result from my theory.
\ Another convenience is that, as demonstrated in \textsc{appendix C} of
\textsc{part II}, it is quite easy to obtain values for the diffusion
parameter $D$, especially when compared to the task of measuring the three
transport parameters required for the NSF formulation.\footnote{One of these
parameters, the Navier-Stokes bulk viscosity, is usually very difficult to
find tabulated.}

Despite these appealing features, there are two aspects of my formulation that
may appear troubling:\ (1) the mechanical mass density is not a directly
measurable quantity, and\ (2) my formulation is incompatible with the kinetic
theory of gases. \ Regarding the first item, it should be observed that there
are many instances of well-accepted macroscopic theories that involve
computationally useful, but not directly measurable, quantities.\ \ The
admitted difference here is that one is accustomed to conceiving of the mass
density as an absolute concept with no ambiguity built into its classical
definition and, therefore, seeming to present no need for alternate
definitions. \ It is typical for more licence to be granted when defining new
velocities and energies, for example. \ Although the mechanical mass density
is an abstract concept, my theory uses it not in place of--but rather in
conjunction with--the actual mass density, which is a quantity that is
measurable and tied, in an averaged sense, to the masses of the molecules
comprising the fluid.\footnote{The physical idea of an impenetrable surface,
for example, requires that no \textit{actual} mass may go through the surface,
whereas this condition does not, in itself, bar \textit{mechanical} mass from
going through the surface.} \ The second of the concerns mentioned above
carries more significance. \ The incompatibility is evident from the fact that
the NSF equations may be derived from the Boltzmann equation via the
Chapman-Enskog procedure for the special case of a classical monatomic ideal
gas (see Huang \cite[\textsc{Ch.}6]{huang}, for example). \ With this
procedure, the NSF formulation is the first-order approximation in the
parameter $K\!n$. \ It must be emphasized that the difference between
Navier-Stokes-Fourier and my theory is \textit{not} a higher-order
$K\!n$\ effect. \ Therefore, mine may not be developed from the Burnett
equations or Grad's moment method, for example. \ The connection between NSF
and kinetic gas theory is generally interpreted to mean that both are correct,
but perhaps this agreement is simply because both are lacking the same
feature, i.e. a mechanism for macroscopic diffusion of mass. \ I realize that
without proof, this seems like a heretical claim. \ However, if the results of
the aforementioned light scattering experiment support my theory, then this
would necessitate a reconception of kinetic gas theory in the hydrodynamic regime.

I have not been alone in my efforts to include an extra diffusive mechanism as
part of the fluid equations of motion. \ For early examples, see Slezkin
\cite{slez1}, \cite{slez2} and Vallander \cite{vall}, and for later examples,
see Klimontovich \cite{klimart}, \cite{klim}, Brenner \cite{bren1},
\cite{bren2}, \cite{bren3}, \cite{bren4}, and Dadzie et al. \cite{dadzie}%
.\footnote{This is not a complete list of references. \ However, it includes
all the primary authors of whom I am aware.} \ My formulation proves to be
significantly different from each of these treatments.

\section{\label{secnt}Notation}

To facilitate the subsequent development, let us first mention a few
notational items. \ The number of lines under a symbol indicates its tensor
order: \ a scalar (zeroth-order tensor) has no underlines (e.g. temperature,
$T$), a first-order tensor has one underline (e.g. velocity, $\underline{v}$),
a second-order tensor has two underlines (e.g. pressure tensor, $\underline
{\underline{P}}$), etc. \ A tensor of arbitrary order is indicated as
$\underline{A}_{\ldots}$. \ The symbol $\underline{\underline{1}}$ is used to
represent the second-order identity tensor and $\underline{0}_{\ldots}$\ to
denote the zero tensor. All of the tensor operators used in this paper are
defined in \textsc{appendix \ref{appto}}.

Extensive quantities, i.e. ones that are additive over composite subsystems,
are denoted using capital letters. \ Below is a list of the extensive
quantities considered here.%
\[%
\begin{array}
[c]{lll}%
\overline{M}=\text{mechanical mass} & \underline{L}=\text{angular momentum} &
W=\text{potential energy}\\
M=\text{mass} & E=\text{total energy} & S=\text{entropy}\\
V=\text{volume} & U=\text{internal energy} & \\
\underline{P}=\text{linear momentum} & K=\text{kinetic energy} &
\end{array}
\]
The densities (amounts per volume) corresponding to the extensive quantities
above are denoted using lower-case letters. \ For example, $\underline{l}$\ is
the angular momentum density, $\overline{m}$\ is the mechanical mass density,
etc. \ For any extensive quantity, $\underline{A}_{\ldots}$, its amount\ per
unit mechanical mass is denoted\ as $\left(  \underline{a}_{\ldots}\right)
_{\overline{M}}$. \ For example, $u_{\overline{M}}$ is the internal energy per
mechanical mass. \ Note that%
\begin{equation}
v_{\overline{M}}=\frac{1}{\overline{m}}\label{nt1}%
\end{equation}
and%
\begin{equation}
\underline{a}_{\ldots}=\overline{m}\left(  \underline{a}_{\ldots}\right)
_{\overline{M}}.\label{nt2}%
\end{equation}

The intensive quantities considered in this paper are%
\[%
\begin{array}
[c]{l}%
T=\text{absolute temperature}\\
P=\text{thermodynamic pressure}\\
\mu=\text{chemical potential (per mass),}%
\end{array}
\]
and the remaining symbols to be defined are%
\[%
\begin{array}
[c]{l}%
\underline{x}=\text{position}\\
t=\text{time}\\
\underline{v}=\text{continuum mechanical velocity}%
\end{array}
\]
and a few others that are defined as they arise in the text.

\section{\label{secbl}Balance Laws}

The general balance law for an extensive quantity, $\underline{A}_{\ldots}$,
is given by%
\begin{equation}
\frac{\partial\underline{a}_{\ldots}}{\partial t}=-\nabla\cdot\left(
\underline{j}_{\ldots}\right)  _{\underline{A}_{\ldots}}+\left(  \underline
{r}_{\ldots}\right)  _{\underline{A}_{\ldots}},\label{bl.1}%
\end{equation}
where $\underline{a}_{\ldots}$\ is the local $\underline{A}_{\ldots}$-density
which is assumed to depend on $\underline{x}$\ and $t$, $\left(  \underline
{j}_{\ldots}\right)  _{\underline{A}_{\ldots}}$\ is the total $\underline
{A}_{\ldots}$-flux, and $\left(  \underline{r}_{\ldots}\right)  _{\underline
{A}_{\ldots}}$\ denotes the volumetric $\underline{A}_{\ldots}$-production/
destruction rate. \ Let us decompose the total flux into its non-convective
and convective parts:%
\begin{equation}
\left(  \underline{j}_{\ldots}\right)  _{\underline{A}_{\ldots}}=\left(
\underline{q}_{\ldots}\right)  _{\underline{A}_{\ldots}}+\underline
{v}\underline{a}_{\ldots}.\label{bl.2}%
\end{equation}

In my proposed continuum formulation for a single-component fluid, I begin by
assuming the local balance laws for the mechanical mass, mass, momentum, total
energy, and entropy to be given respectively by%
\begin{equation}
\frac{\partial\overline{m}}{\partial t}=-\nabla\cdot\left(  \overline
{m}\underline{v}\right) \label{bl1}%
\end{equation}%
\begin{equation}
\frac{\partial m}{\partial t}=-\nabla\cdot\left(  \underline{q}_{M}%
+m\underline{v}\right) \label{bl2}%
\end{equation}%
\begin{equation}
\frac{\partial\left(  \overline{m}\underline{v}\right)  }{\partial t}%
=-\nabla\cdot\left(  \underline{\underline{P}}+\overline{m}\underline
{v}\,\underline{v}\right)  +\overline{m}\underline{f}_{\overline{M}%
}\label{bl3}%
\end{equation}%
\begin{equation}
\frac{\partial e}{\partial t}=-\nabla\cdot\left(  \underline{q}_{\text{heat}%
}+\underline{q}_{\text{chem}}+\underline{\underline{P}}\cdot\underline
{v}+e\underline{v}\right) \label{bl4}%
\end{equation}
and%
\begin{equation}
\frac{\partial s}{\partial t}=-\nabla\cdot\left(  \underline{q}_{S}%
+s\underline{v}\right)  +r_{S}.\label{bl5}%
\end{equation}
Note that equation (\ref{bl1}) is a continuity equation for the mechanical
mass density and that equation (\ref{bl2}) governing the actual mass density
contains an additional non-convective mass flux. \ In momentum equation
(\ref{bl3}), the momentum density is assumed to be%
\begin{equation}
\underline{p}=\overline{m}\underline{v}.\label{bl5.5}%
\end{equation}
Also, $\underline{\underline{P}}$ denotes the total pressure tensor, i.e. the
non-convective momentum flux, $\underline{\underline{q}}_{\underline{P}}$, and
$\underline{f}_{\overline{M}}$ is defined to be the external body force per
mechanical mass, assumed conservative, which in this setting means to satisfy%
\begin{equation}
\underline{f}_{\overline{M}}=-\nabla w_{\overline{M}}\label{bl6}%
\end{equation}
with%
\begin{equation}
\frac{\partial w_{\overline{M}}}{\partial t}=0.\label{bl6.5}%
\end{equation}
In the total energy equation (\ref{bl4}), there appear, as usual, the heat
flux, $\underline{q}_{\text{heat}}$, and the energy flux arising from
mechanical work, $\underline{\underline{P}}\cdot\underline{v}$, but I have
also included an energy flux due to chemical work, $\underline{q}%
_{\text{chem}}$.

Based on equations (\ref{bl1})-(\ref{bl5}), there are a few others which may
be derived. \ Defining the kinetic energy density as%
\begin{equation}
k=\frac{1}{2}\overline{m}\left\Vert \underline{v}\right\Vert ^{2},\label{bl7}%
\end{equation}
it is shown as computation (1) in \textsc{appendix \ref{appto}}\ that
equations (\ref{bl1}) and (\ref{bl3}) may be used along with a few tensor
identities to compute the following kinetic energy balance law:%
\begin{equation}
\frac{\partial k}{\partial t}=-\nabla\cdot\left(  \underline{\underline{P}%
}\cdot\underline{v}+k\underline{v}\right)  +\underline{\underline{P}}%
^{T}\colon\nabla\underline{v}+\overline{m}\underline{v}\cdot\underline
{f}_{\overline{M}},\label{bl7.5}%
\end{equation}
or using (\ref{bl6}) in the last term,%
\begin{equation}
\frac{\partial k}{\partial t}=-\nabla\cdot\left(  \underline{\underline{P}%
}\cdot\underline{v}+k\underline{v}\right)  +\underline{\underline{P}}%
^{T}\colon\nabla\underline{v}-\overline{m}\underline{v}\cdot\nabla
w_{\overline{M}}.\label{bl8}%
\end{equation}
Also, since (\ref{nt2}) implies that the potential energy density may be
expressed as%
\begin{equation}
w=\overline{m}w_{\overline{M}},\label{bl8.5}%
\end{equation}
one may employ (\ref{bl1}) and assumption (\ref{bl6.5}), to obtain the
following equation for the potential energy:%
\begin{equation}
\frac{\partial w}{\partial t}=-w_{\overline{M}}\nabla\cdot\left(  \overline
{m}\underline{v}\right)  .\label{bl9}%
\end{equation}
Therefore, by defining the internal energy density, $u$, via%
\begin{equation}
e=u+k+w,\label{bl10}%
\end{equation}
one finds that (\ref{bl4}), (\ref{bl8}), and (\ref{bl9}), together with
identity (\ref{to25}), yield the internal energy balance law,%
\begin{equation}
\frac{\partial u}{\partial t}=-\nabla\cdot\left(  \underline{q}_{\text{heat}%
}+\underline{q}_{\text{chem}}+u\underline{v}\right)  -\underline{\underline
{P}}^{T}\colon\nabla\underline{v}.\label{bl11}%
\end{equation}
In view of the above, the non-convective internal energy flux is given by%
\begin{equation}
\underline{q}_{U}=\underline{q}_{\text{heat}}+\underline{q}_{\text{chem}%
}.\label{bl12}%
\end{equation}

Next, let us define the angular momentum density to be%
\begin{equation}
\underline{l}=\underline{x}\times\left(  \overline{m}\underline{v}\right)
\label{bl12.5}%
\end{equation}
and use (\ref{bl3}) to compute the angular momentum balance as%
\begin{align}
\frac{\partial\left[  \underline{x}\times\left(  \overline{m}\underline
{v}\right)  \right]  }{\partial t} &  =\underline{x}\times\frac{\partial
\left(  \overline{m}\underline{v}\right)  }{\partial t}\nonumber\\
&  =-\underline{x}\times\left[  \nabla\cdot\left(  \underline{\underline{P}%
}+\overline{m}\underline{v}\,\underline{v}\right)  \right]  +\underline
{x}\times\left(  \overline{m}\underline{f}_{\overline{M}}\right)
.\label{bl13}%
\end{align}
For the angular momentum to be conserved,%
\begin{equation}
\frac{\partial\left[  \underline{x}\times\left(  \overline{m}\underline
{v}\right)  \right]  }{\partial t}=-\nabla\cdot\left[  \underline{x}%
\times\left(  \underline{\underline{P}}+\overline{m}\underline{v}%
\,\underline{v}\right)  \right]  +\underline{x}\times\left(  \overline
{m}\underline{f}_{\overline{M}}\right) \label{bl14}%
\end{equation}
must be satisfied, and as computation (2) in \textsc{appendix \ref{appto}},
this is demonstrated to be true if and only if the pressure tensor is
symmetric:%
\begin{equation}
\underline{\underline{P}}=\underline{\underline{P}}^{T}.\label{bl15}%
\end{equation}
Henceforth, let us assume (\ref{bl15}) to hold.

In preparation for the next section, definition (\ref{to21}) of the convective
derivative may be used to rewrite some of these balance laws. \ Employing the
relation (\ref{nt1}) and identity (\ref{to25}), one may write (\ref{bl1}) as%
\begin{equation}
\overline{m}\frac{Dv_{\overline{M}}}{Dt}=\nabla\cdot\underline{v}.\label{bl17}%
\end{equation}
Also, using (\ref{nt2}), (\ref{bl1}), and (\ref{to25}), one may express
equations (\ref{bl2}), (\ref{bl11}), and (\ref{bl5}) as%
\begin{equation}
\overline{m}\frac{Dm_{\overline{M}}}{Dt}=-\nabla\cdot\underline{q}%
_{M},\label{bl18}%
\end{equation}%
\begin{equation}
\overline{m}\frac{Du_{\overline{M}}}{Dt}=-\nabla\cdot\left(  \underline
{q}_{\text{heat}}+\underline{q}_{\text{chem}}\right)  -\underline
{\underline{P}}\colon\nabla\underline{v},\label{bl19}%
\end{equation}
and%
\begin{equation}
\overline{m}\frac{Ds_{\overline{M}}}{Dt}=-\nabla\cdot\underline{q}_{S}%
+r_{S},\label{bl20}%
\end{equation}
where (\ref{bl15}) has been assumed in equation (\ref{bl19}).

In order to compare the foregoing balance laws with those typically postulated
when deriving the Navier-Stokes-Fourier formulation, see de Groot and Mazur
\cite[\textsc{ch}. \textsc{II}]{degroot}, for example.

\section{\label{secce}Constitutive Equations}

Let us study phenomena close enough to equilibrium so that classical
irreversible thermodynamics is applicable. \ Below, this formalism, which is
detailed in de Groot and Mazur \cite[\textsc{ch}. \textsc{III} and
\textsc{IV}]{degroot}, is used to obtain linear non-convective constitutive
equations for the fluxes.

For a single-component fluid, I propose an equilibrium fundamental relation
for the entropy per mechanical mass of the form,%
\begin{equation}
\widehat{s}_{\overline{M}}=\widetilde{s}_{\overline{M}}\left(  \widehat
{u}_{\overline{M}},\widehat{v}_{\overline{M}},\widehat{m}_{\overline{M}%
}\right)  .\label{ce1}%
\end{equation}
Note the following equilibrium thermodynamic relationships:%
\begin{equation}
\frac{1}{\widehat{T}}=\frac{\partial\widetilde{s}_{\overline{M}}}%
{\partial\widehat{u}_{\overline{M}}}\text{, }\frac{\widehat{P}}{\widehat{T}%
}=\frac{\partial\widetilde{s}_{\overline{M}}}{\partial\widehat{v}%
_{\overline{M}}}\text{, and }\frac{\widehat{\mu}}{\widehat{T}}=-\frac
{\partial\widetilde{s}_{\overline{M}}}{\partial\widehat{m}_{\overline{M}}%
}.\label{ce2}%
\end{equation}

Let us assume the local equilibrium hypothesis, which implies that if
$\widehat{\alpha}$\ represents any of the thermodynamic parameters mentioned
above, then its corresponding local variable can be defined as%
\begin{equation}
\alpha=\left.  \widehat{\alpha}\right\vert _{\left(  \underline{x},t\right)
}.\label{ce5}%
\end{equation}
In particular, the fundamental equation (\ref{ce1}) may be written locally as%
\begin{equation}
s_{\overline{M}}=\widetilde{s}_{\overline{M}}\left(  u_{\overline{M}%
},v_{\overline{M}},m_{\overline{M}}\right) \label{ce6}%
\end{equation}
which, upon taking its total differential, yields%
\begin{equation}
ds_{\overline{M}}=\frac{1}{T}du_{\overline{M}}+\frac{P}{T}dv_{\overline{M}%
}-\frac{\mu}{T}dm_{\overline{M}},\label{ce7}%
\end{equation}
provided that the differentiation is performed in a reference frame for which
$\overline{M}$\ is constant. \ Therefore, we may take the convective
derivative of the above\ and multiply through by $\overline{m}$ to obtain%
\begin{equation}
\overline{m}\frac{Ds_{\overline{M}}}{Dt}=\frac{1}{T}\overline{m}%
\frac{Du_{\overline{M}}}{Dt}+\frac{P}{T}\overline{m}\frac{Dv_{\overline{M}}%
}{Dt}-\frac{\mu}{T}\overline{m}\frac{Dm_{\overline{M}}}{Dt}.\label{ce8}%
\end{equation}
Substituting equations (\ref{bl17})-(\ref{bl20}) into the above and using
identity (\ref{to28}), one arrives at the following expression for the
volumetric rate of entropy production:%
\begin{equation}
r_{S}=\nabla\cdot\underline{q}_{S}-\frac{1}{T}\nabla\cdot\left(  \underline
{q}_{\text{heat}}+\underline{q}_{\text{chem}}\right)  -\frac{1}{T}\left(
\underline{\underline{P}}-P\underline{\underline{1}}\right)  \colon
\nabla\underline{v}+\frac{\mu}{T}\nabla\cdot\underline{q}_{M}.\label{ce9}%
\end{equation}

In the differential form\ of the first law of equilibrium thermodynamics, the
change in energy due to heat and chemical work are represented by $TdS$ and
$\mu dM$, respectively. \ With this in mind, let us take%
\begin{equation}
\underline{q}_{\text{heat}}=T\underline{q}_{S}\label{ce9.1}%
\end{equation}
and%
\begin{equation}
\underline{q}_{\text{chem}}=\mu\underline{q}_{M}.\label{ce9.2}%
\end{equation}
By employing these in equation (\ref{ce9}) and using (\ref{bl12}) and
(\ref{to25}), one finds%
\begin{equation}
r_{S}=\underline{q}_{U}\cdot\nabla\left(  \frac{1}{T}\right)  -\frac{1}%
{T}\left(  \underline{\underline{P}}-P\underline{\underline{I}}\right)
\colon\nabla\underline{v}-\underline{q}_{M}\cdot\nabla\left(  \frac{\mu}%
{T}\right)  .\label{ce9.3}%
\end{equation}

As discussed in \textsc{appendix \ref{applc}}, the above expression for the
entropy production rate motivates the following linear constitutive equations:%
\begin{equation}
\underline{q}_{M}=-d_{M}\nabla\mu-\frac{k_{M}}{\mu}\nabla T,\label{ce12}%
\end{equation}%
\begin{equation}
\underline{\underline{P}}=P\underline{\underline{1}}+\underline{\underline{P}%
}_{\text{visc}}\text{ with }\underline{\underline{P}}_{\text{visc}}%
=-\zeta\left(  \nabla\cdot\underline{v}\right)  \underline{\underline{1}%
}-2\eta\left(  \nabla\underline{v}\right)  ^{sy,dev},\label{ce13}%
\end{equation}
and%
\begin{equation}
\underline{q}_{U}=-k_{U}\nabla T-\mu d_{U}\nabla\mu,\label{ce14}%
\end{equation}
and Onsager reciprocity requires the coefficients to satisfy the relation,%
\begin{equation}
\mu\left(  d_{U}-d_{M}\right)  =\frac{T}{\mu}k_{M}.\label{ce16}%
\end{equation}

\section{Summary of the $\overline{M}$-Formulation}

My general formulation, which I refer to as the $\overline{M}$-formulation,
may be written as equations (\ref{bl1})-(\ref{bl3}) and (\ref{bl11}%
)/(\ref{bl12}),%
\begin{align}
\frac{\partial\overline{m}}{\partial t}  & =-\nabla\cdot\left(  \overline
{m}\underline{v}\right) \nonumber\\
\frac{\partial m}{\partial t}  & =-\nabla\cdot\left(  \underline{q}%
_{M}+m\underline{v}\right) \label{smf1}\\
\frac{\partial\left(  \overline{m}\underline{v}\right)  }{\partial t}  &
=-\nabla\cdot\left(  \underline{\underline{P}}+\overline{m}\underline
{v}\,\underline{v}\right)  +\overline{m}\underline{f}_{\overline{M}%
}\nonumber\\
\frac{\partial u}{\partial t}  & =-\nabla\cdot\left(  \underline{q}%
_{U}+u\underline{v}\right)  -\underline{\underline{P}}\colon\nabla
\underline{v}\nonumber
\end{align}
with constitutive equations (\ref{ce12})-(\ref{ce14}),%
\begin{align}
\underline{q}_{M}  & =-d_{M}\nabla\mu-\frac{k_{M}}{\mu}\nabla T\nonumber\\
\underline{\underline{P}}  & =P\underline{\underline{1}}+\underline
{\underline{P}}_{\text{visc}}\text{ with }\underline{\underline{P}%
}_{\text{visc}}=-\zeta\left(  \nabla\cdot\underline{v}\right)  \underline
{\underline{1}}-2\eta\left(  \nabla\underline{v}\right)  ^{sy,dev}%
\label{smf2}\\
\underline{q}_{U}  & =-k_{U}\nabla T-\mu d_{U}\nabla\mu,\nonumber
\end{align}
where the transport coefficients, $d_{M}$, $k_{M}$, and $d_{U}$,\ are assumed
to satisfy (\ref{ce16}). \ The $\overline{M}$-formulation may be shown to obey
Galilean invariance and the conservation of angular momentum. \ Furthermore,
the constitutive laws each satisfy the principle of objectivity required for
material frame indifference. \ That is to say, all of the fluxes in
(\ref{smf2}) transform in the expected way under time-dependent rigid body
translation and rotation.

Using the notation from \S \ref{secnt}, the Navier-Stokes-Fourier formulation
may be written as%
\begin{align}
\frac{\partial m}{\partial t}  & =-\nabla\cdot\left(  m\underline{v}\right)
\nonumber\\
\frac{\partial\left(  m\underline{v}\right)  }{\partial t}  & =-\nabla
\cdot\left(  \underline{\underline{P}}+m\underline{v}\,\underline{v}\right)
+m\underline{f}_{M}\label{smf3}\\
\frac{\partial u}{\partial t}  & =-\nabla\cdot\left(  \underline{q}%
_{U}+u\underline{v}\right)  -\underline{\underline{P}}\colon\nabla
\underline{v}\nonumber
\end{align}
with constitutive equations,%
\begin{align}
\underline{\underline{P}}  & =P\underline{\underline{I}}+\underline
{\underline{P}}_{\text{visc}}\text{ with }\underline{\underline{P}%
}_{\text{visc}}=-\zeta_{NS}\left(  \nabla\cdot\underline{v}\right)
\underline{\underline{I}}-2\eta_{NS}\left(  \nabla\underline{v}\right)
^{sy,dev}\label{smf4}\\
\underline{q}_{U}  & =-k_{F}\nabla T,\nonumber
\end{align}
where $k_{F}$\ is the Fourier heat conductivity, $\zeta_{NS}$\ and $\eta_{NS}%
$\ are the Navier-Stokes bulk and shear viscosities, respectively, and
$\underline{f}_{M}$\ represents the specific body force. \ It is clear that if
one chooses%
\begin{gather}
d_{M}=k_{M}=d_{U}=0\nonumber\\
k_{U}=k_{F}\label{smf4.1}\\
\zeta=\zeta_{NS}\nonumber\\
\eta=\eta_{NS}\nonumber
\end{gather}
in the $\overline{M}$-formulation, then $m$\ and $\overline{m}$\ are governed
by the same differential equation. \ If, in addition, initial and boundary
conditions are taken to be the same for $\overline{m}$ as for $m$, then these
two variables are equal, that is%
\begin{equation}
\overline{m}=m,\label{smf4.2}%
\end{equation}
resulting in the NSF formulation above. \ Finally, note that choices
(\ref{smf4.1} a) satisfy Onsager reciprocity condition (\ref{ce16}).
\ Therefore, the NSF formulation may be considered as a special case of the
general $\overline{M}$-formulation.

Next, let us employ the equilibrium thermodynamic relations (\ref{e1}%
)-(\ref{e3}) to express the general $\overline{M}$-formulation (\ref{smf1}%
)/(\ref{smf2}) entirely in terms of the dependent variables, $\overline{m}$,
$m$, $\underline{v}$, and $u$. \ Doing so, one obtains%
\begin{align}
\frac{\partial\overline{m}}{\partial t}  & =-\nabla\cdot\left(  \overline
{m}\underline{v}\right) \nonumber\\
\frac{\partial m}{\partial t}  & =-\nabla\cdot\left(  \underline{q}%
_{M}+m\underline{v}\right) \label{smf5}\\
\frac{\partial\left(  \overline{m}\underline{v}\right)  }{\partial t}  &
=\beta_{m}\nabla m+\beta_{u}\nabla u-\nabla\cdot\left(  \underline
{\underline{P}}_{\text{visc}}+\overline{m}\underline{v}\,\underline{v}\right)
+\overline{m}\underline{f}_{\overline{M}}\nonumber\\
\frac{\partial u}{\partial t}  & =-\nabla\cdot\left(  \underline{q}%
_{U}+u\underline{v}\right)  -P\left(  m,u\right)  \nabla\cdot\underline
{v}-\underline{\underline{P}}_{\text{visc}}\colon\nabla\underline{v}\nonumber
\end{align}
with%
\begin{align}
\underline{q}_{M}  & =-\nu_{m}\nabla m-\nu_{mu}\nabla u\nonumber\\
\underline{\underline{P}}_{\text{visc}}  & =-\zeta\left(  \nabla
\cdot\underline{v}\right)  \underline{\underline{1}}-2\eta\left(
\nabla\underline{v}\right)  ^{sy,dev}\label{smf6}\\
\underline{q}_{U}  & =-\nu_{um}\nabla m-\nu_{u}\nabla u,\nonumber
\end{align}
where the following coefficients have been defined:%
\begin{align}
\beta_{m}  & =\frac{\alpha_{P}h_{M}-\frac{T\alpha_{P}^{2}}{m\kappa_{T}}-c_{V}%
}{m\kappa_{T}c_{V}}\label{smf7}\\
\beta_{u}  & =-\frac{\alpha_{P}}{m\kappa_{T}c_{V}},\label{smf8}%
\end{align}%
\begin{align}
\nu_{m}  & =\left\{
\begin{array}
[c]{c}%
\left[  \frac{1}{m^{2}\kappa_{T}}+\frac{\left(  h_{M}-\frac{T\alpha_{P}%
}{m\kappa_{T}}\right)  \left(  h_{M}-\frac{T\alpha_{P}}{m\kappa_{T}}%
-\mu\right)  }{mTc_{V}}\right]  d_{M}-\\
\frac{\left(  h_{M}-\frac{T\alpha_{P}}{m\kappa_{T}}\right)  }{mc_{V}\mu}k_{M}%
\end{array}
\right\} \label{smf9}\\
\nu_{mu}  & =-\frac{\left(  h_{M}-\frac{T\alpha_{P}}{m\kappa_{T}}-\mu\right)
}{mTc_{V}}d_{M}+\frac{1}{mc_{V}\mu}k_{M}\label{smf10}\\
\nu_{um}  & =\left\{
\begin{array}
[c]{c}%
-\frac{\left(  h_{M}-\frac{T\alpha_{P}}{m\kappa_{T}}\right)  }{mc_{V}}k_{U}+\\
\left[  \frac{\mu}{m^{2}\kappa_{T}}+\frac{\mu\left(  h_{M}-\frac{T\alpha_{P}%
}{m\kappa_{T}}\right)  \left(  h_{M}-\frac{T\alpha_{P}}{m\kappa_{T}}%
-\mu\right)  }{mTc_{V}}\right]  d_{U}%
\end{array}
\right\} \label{smf11}\\
\nu_{u}  & =\frac{1}{mc_{V}}k_{U}-\frac{\mu\left(  h_{M}-\frac{T\alpha_{P}%
}{m\kappa_{T}}-\mu\right)  }{mTc_{V}}d_{U}\label{smf12}%
\end{align}
with specific enthalpy $h_{M}$, isothermal compressibility $\kappa_{T}$,
thermal expansion coefficient $\alpha_{P}$,\ and isochoric specific heat per
mass $c_{V}$, and I have indicated that the thermodynamic pressure, $P$,
should be written in terms of the variables, $u$\ and $m$, e.g. to model a
classical monatomic ideal gas, one takes%
\begin{equation}
P\left(  m,u\right)  =\frac{2}{3}u.\label{smf12.3}%
\end{equation}

\section{\label{seclin}Linearization}

If one assumes there to be a constant equilibrium state,%
\begin{equation}
\left(  \overline{m},m,\underline{v},u\right)  =\left(  m_{\text{eq}%
},m_{\text{eq}},\underline{0},u_{\text{eq}}\right)  ,\label{smf13}%
\end{equation}
and that the variables do not deviate too far from this state, then one finds
the linearization of (\ref{smf1})/(\ref{smf2}), given by%
\begin{align}
\frac{\partial\overline{m}}{\partial t}  & =\nabla\cdot\underline
{p}\nonumber\\
\frac{\partial m}{\partial t}  & =\left(  \nu_{m}\right)  _{\text{eq}}%
\nabla^{2}m+\left(  \nu_{mu}\right)  _{\text{eq}}\nabla^{2}u-\nabla
\cdot\underline{p}\label{smf14}\\
\frac{\partial\underline{p}}{\partial t}  & =\left[
\begin{array}
[c]{c}%
\left(  \beta_{m}\right)  _{\text{eq}}\nabla m+\left(  \beta_{u}\right)
_{\text{eq}}\nabla u+\\
\left(  \nu_{R}\right)  _{\text{eq}}\nabla^{2}\underline{p}+\left(  \nu
_{L}-\nu_{R}\right)  _{\text{eq}}\nabla\left(  \nabla\cdot\underline
{p}\right)
\end{array}
\right] \nonumber\\
\frac{\partial u}{\partial t}  & =\left(  \nu_{um}\right)  _{\text{eq}}%
\nabla^{2}m+\left(  \nu_{u}\right)  _{\text{eq}}\nabla^{2}u-\left(
h_{M}\right)  _{\text{eq}}\nabla\cdot\underline{p}\nonumber
\end{align}
to be a good approximate description. \ In the above, subscript "eq" denotes
evaluation at the equilibrium state (\ref{smf13}), $\underline{p}$\ is the
momentum defined by (\ref{bl5.5}) and approximated near equilibrium as%
\begin{equation}
\underline{p}=m_{\text{eq}}\underline{v},\label{smf14.3}%
\end{equation}
and external body forces have been omitted. \ Furthermore, I have used
equilibrium thermodynamic relationship (\ref{e10}), tensor identities
(\ref{to25}) and (\ref{to31}), and the following definitions of the
longitudinal and rotational kinematic viscosities:%
\begin{equation}
\nu_{L}=\frac{\left(  \zeta+\frac{4}{3}\eta\right)  }{\overline{m}}\text{ and
}\nu_{R}=\frac{\eta}{\overline{m}}.\label{smf15}%
\end{equation}
The above kinematic viscosities are so-named because if the linearized
momentum is decomposed into its longitudinal and rotational parts via%
\begin{equation}
\underline{p}=\underline{p}_{L}+\underline{p}_{R}\label{smf16}%
\end{equation}
with%
\begin{equation}
\nabla\times\underline{p}_{L}=\underline{0}\text{ and }\nabla\cdot
\underline{p}_{R}=0,\label{smf17}%
\end{equation}
then one may use tensor identity (\ref{to32}) to express (\ref{smf14}) as a
longitudinal part,%
\begin{align}
\frac{\partial\overline{m}}{\partial t}  & =\nabla\cdot\underline{p}%
_{L}\nonumber\\
\frac{\partial m}{\partial t}  & =\left(  \nu_{m}\right)  _{\text{eq}}%
\nabla^{2}m+\left(  \nu_{mu}\right)  _{\text{eq}}\nabla^{2}u-\nabla
\cdot\underline{p}_{L}\label{smf18}\\
\frac{\partial\underline{p}_{L}}{\partial t}  & =\left(  \beta_{m}\right)
_{\text{eq}}\nabla m+\left(  \beta_{u}\right)  _{\text{eq}}\nabla u+\left(
\nu_{L}\right)  _{\text{eq}}\nabla^{2}\underline{p}_{L}\nonumber\\
\frac{\partial u}{\partial t}  & =\left(  \nu_{um}\right)  _{\text{eq}}%
\nabla^{2}m+\left(  \nu_{u}\right)  _{\text{eq}}\nabla^{2}u-\left(
h_{M}\right)  _{\text{eq}}\nabla\cdot\underline{p}_{L}\nonumber
\end{align}
and a rotational part,%
\begin{equation}
\frac{\partial\underline{p}_{R}}{\partial t}=\left(  \nu_{R}\right)
_{\text{eq}}\nabla^{2}\underline{p}_{R},\label{smf19}%
\end{equation}
that are independent of one another, provided that boundary conditions do not
couple them together.

\section{\label{sectc}Transport Coefficients}

First, let us note that choosing (\ref{smf4.1}) and (\ref{smf4.2}) in
definitions (\ref{smf9})-(\ref{smf12}) and (\ref{smf15}) yields the following
coefficients for the Navier-Stokes-Fourier formulation:%
\begin{gather}
\nu_{m}=\nu_{mu}=0\nonumber\\
\nu_{um}=-\frac{\left(  h_{M}-\frac{T\alpha_{P}}{m\kappa_{T}}\right)  }%
{mc_{V}}k_{F}\text{ and }\nu_{u}=\frac{1}{mc_{V}}k_{F}\label{tc0}\\
\nu_{L}=\frac{\left(  \zeta_{NS}+\frac{4}{3}\eta_{NS}\right)  }{m}\text{ and
}\nu_{R}=\frac{\eta_{NS}}{m}.\nonumber
\end{gather}

On the other hand, in view of decomposition (\ref{smf18})/(\ref{smf19}), I
propose choosing these coefficients in a different way. By assuming the
dissipative mechanism associated with the longitudinal quantities, i.e. the
mass, internal energy, and longitudinal momentum, is diffusion, all
characterized by the same diffusion coefficient $D$, one takes%
\begin{gather}
\nu_{m}=\nu_{u}=\nu_{L}=D\label{tc.1}\\
\nu_{mu}=\nu_{um}=0.\label{tc.11}%
\end{gather}
In addition, since the NSF formulation and my formulation both lead to the
same rotational momentum equation (\ref{smf19}), which is independent of the
rest of the equations, I postulate that the shear viscosity is the same for
both descriptions:%
\begin{equation}
\eta=\eta_{NS}.\label{tc.2}%
\end{equation}
Note that (\ref{smf19}) implies the dissipative mechanism associated with the
rotational momentum is--as for the longitudinal quantities--diffusion, but
with its own characteristic diffusion coefficient, $\nu_{R}$. \ I refer to the
$\overline{M}$-formulation, with choices (\ref{tc.1})-(\ref{tc.2}) for the
transport coefficients, as the $\overline{M}\left(  D,\eta\right)  $-formulation.

If one substitutes (\ref{tc.1}) and (\ref{tc.11}) into expressions
(\ref{smf9})-(\ref{smf12}) and solves for the transport parameters appearing
in constitutive equations (\ref{ce12}) and (\ref{ce14}), then one finds%
\begin{align}
d_{M}  & =m^{2}\kappa_{T}D\nonumber\\
k_{M}  & =\frac{m^{2}\mu\kappa_{T}}{T}\left(  h_{M}-\mu-\frac{T\alpha_{P}%
}{m\kappa_{T}}\right)  D\label{tc.25}\\
d_{U}  & =\frac{m^{2}\kappa_{T}}{\mu}\left(  h_{M}-\frac{T\alpha_{P}}%
{m\kappa_{T}}\right)  D\nonumber\\
k_{U}  & =m\left[  c_{V}+\frac{m\kappa_{T}}{T}\left(  h_{M}-\frac{T\alpha_{P}%
}{m\kappa_{T}}\right)  \left(  h_{M}-\frac{T\alpha_{P}}{m\kappa_{T}}%
-\mu\right)  \right]  D,\nonumber
\end{align}
and it may be shown that these satisfy Onsager reciprocity condition
(\ref{ce16}).

Substituting (\ref{tc.1}) into the longitudinal kinematic viscosity definition
(\ref{smf15} a) and solving for the bulk viscosity, gives%
\begin{equation}
\zeta=\overline{m}D-\frac{4}{3}\eta\label{tc.3}%
\end{equation}
for the $\overline{M}\left(  D,\eta\right)  $-formulation. \ We will find this
quantity to be different from its Navier-Stokes counterpart. \ Consequently,
while (\ref{tc.2}) justifies dropping the subscript "$NS$" and simply
referring to $\eta$\ as the shear viscosity, we may not treat the bulk
viscosity in the same manner, and thus I will continue to refer to the
$\overline{M}\left(  D,\eta\right)  $-formulation bulk viscosity as $\zeta
$\ and the Navier-Stokes bulk viscosity as $\zeta_{NS}$.

For convenience, let us define a dimensionless coefficient $C$\ (possibly a
function of thermodynamic variables) via%
\begin{equation}
\overline{m}D=C\eta.\label{tc22}%
\end{equation}
In order to match Greenspan's \cite{green56} sound attenuation data in the
hydrodynamic regime for classical monatomic ideal gases at room temperature,
for example, we take $C=7/6$. \ (See \S 5 of \textsc{part II}.) \ Using
(\ref{tc22}) in (\ref{tc.3}) yields the following expression for the
$\overline{M}\left(  D,\eta\right)  $-formulation bulk viscosity:%
\begin{equation}
\zeta=\left(  C-\frac{4}{3}\right)  \eta.\label{tc.4}%
\end{equation}
Note that this gives the negatively-valued, $\zeta=-\eta/6$,\ for the
monatomic gas mentioned above, whereas in the NSF formulation for this case,
$\zeta_{NS}$ is taken to be zero.\footnote{If one were to enforce the second
law of thermodynamics as in de Groot and Mazur \cite[\textsc{ch. IV}]{degroot}
by requiring that $r_{S}\geq0$, then a bulk viscosity such that $\zeta<0$
could potentially violate this law. \ However, in a future paper
\cite{stochIII}, I will argue that the second law of thermodynamics should
actually be enforced on a different measure of the entropy production rate,
and that doing so leads to the weaker requirement that $\zeta+4\eta/3\geq0$.}
\ In \textsc{appendix} \textsc{\ C}\ of \textsc{part II}, values of $C$\ and
$D$\ for various types of gases and liquids are computed using ideas from \S 5
of \textsc{part II}, comparisons between $D$\ and the self-diffusion
coefficient $D_{\text{self}}$\ are made, and the temperature and pressure
dependence of $D$\ is examined.

To summarize, the $\overline{M}\left(  D,\eta\right)  $-formulation
constitutive equations are%
\begin{align}
\underline{q}_{M}  & =-D\nabla m\nonumber\\
\underline{\underline{P}}  & =P\underline{\underline{1}}+\underline
{\underline{P}}_{\text{visc}}\text{ with }\underline{\underline{P}%
}_{\text{visc}}=\left[
\begin{array}
[c]{c}%
-\left(  \overline{m}D-\frac{4\eta}{3}\right)  \left(  \nabla\cdot
\underline{v}\right)  \underline{\underline{1}}-\\
2\eta\left(  \nabla\underline{v}\right)  ^{sy,dev}%
\end{array}
\right] \label{tc.5}\\
\underline{q}_{U}  & =-D\nabla u.\nonumber
\end{align}

\section{Conclusions and Future Work}

I hope to have motivated the proposed $\overline{M}\left(  D,\eta\right)
$-formulation by the simple transport assumptions underlying its constitutive
equations, ones that are easily generalized to model more complicated problems
such as multicomponent fluid mixtures (see \textsc{appendix \ref{appmc}}).
\ Of course, its utility nonetheless hinges on the answers to two important
questions:\ (1) how adequately does this formulation perform in problems of
fluid mechanics, and (2) are its predictions in the hydrodynamic regime
quantifiably different from those found with the Navier-Stokes-Fourier
formulation? \ Next in \textsc{part II}, I highlight some of the results I
have obtained so far.

\appendix

\section{\label{appto}Tensors}

In this appendix, indicial notation is used in which sums are taken over
repeated indices. \ The indices may assume the values 1, 2, or 3, representing
the three spatial directions.

\subsubsection*{Algebraic Operators}

\begin{itemize}
\item dots and contraction operators:%
\begin{align}
\underline{u}\cdot\underline{w}  & =u_{i}w_{i}\label{to1}\\
\left(  \underline{\underline{T}}\cdot\underline{w}\right)  _{i}  &
=T_{ij}w_{j}\label{to2}\\
\left(  \underline{w}\cdot\underline{\underline{T}}\right)  _{i}  &
=w_{j}T_{ji}\label{to2.5}%
\end{align}%
\begin{align}
\underline{\underline{T}}\cdot\cdot\underline{\underline{U}}  & =T_{ij}%
U_{ij}\label{to3}\\
\underline{\underline{T}}\colon\underline{\underline{U}}  & =T_{ij}%
U_{ji}\label{to4}%
\end{align}%
\begin{align}
C_{1,2}\left(  \underline{\underline{T}}\right)   & =T_{ii}=\text{tr}\left(
\underline{\underline{T}}\right) \label{to5}\\
\left[  C_{1,3}\left(  \underline{\underline{\underline{D}}}\right)  \right]
_{j}  & =D_{iji}\label{to6}%
\end{align}
If $\underline{A}_{\ldots}$\ and $\underline{B}_{\ldots}$\ are tensors of
order $M$\ and $N$, respectively, with $M\geq N$, then%
\begin{equation}
\left[  \underline{A}_{\ldots}\left(  \cdot\right)  \underline{B}_{\ldots
}\right]  _{i_{1}\ldots i_{M-N}}=A_{i_{1}\ldots i_{M-N}j_{1}\ldots j_{N}%
}B_{j_{1}\ldots j_{N}}.\label{to6.5}%
\end{equation}

\item norms:%
\begin{equation}
\left\Vert \underline{w}\right\Vert =\sqrt{\underline{w}\cdot\underline{w}%
}\text{ and }\left\Vert \underline{\underline{T}}\right\Vert =\sqrt
{\underline{\underline{T}}\cdot\cdot\underline{\underline{T}}}\label{to7}%
\end{equation}

\item tensor product:%
\begin{equation}
\left(  \underline{u}\,\underline{w}\right)  _{ij}=u_{i}w_{j}\label{to8}%
\end{equation}

\item cross products:%
\begin{align}
\left(  \underline{u}\times\underline{w}\right)  _{i}  & =u_{j}w_{k}%
\varepsilon_{ijk}\label{to9}\\
\left(  \underline{w}\times\underline{\underline{T}}\right)  _{ij}  &
=w_{k}T_{il}\varepsilon_{jkl}\label{to10}\\
\left(  \underline{\underline{T}}\times\underline{\underline{U}}\right)
_{ijk}  & =T_{il}U_{jm}\varepsilon_{klm}\label{to11}%
\end{align}
where $\varepsilon_{ijk}$\ is the alternating symbol,%
\begin{equation}
\varepsilon_{ijk}=\left\{
\begin{array}
[c]{l}%
1\text{ if }(i,j,k)=(1,2,3),(2,3,1),\text{ or }(3,1,2)\\
-1\text{ if }(i,j,k)=(1,3,2),(2,1,3),\text{ or }(3,2,1)\\
0\text{ if any of the indices are repeated}%
\end{array}
\right\} \label{to12}%
\end{equation}

\end{itemize}

The transpose, symmetric part, skew-symmetric part, spherical part, and
deviatoric part of a tensor are defined by%
\begin{align}
\left(  \underline{\underline{T}}^{T}\right)  _{ij}  & =T_{ji}\label{to13}\\
\underline{\underline{T}}^{sy}  & =\frac{1}{2}\left(  \underline{\underline
{T}}+\underline{\underline{T}}^{T}\right) \label{to14}\\
\underline{\underline{T}}^{asy}  & =\frac{1}{2}\left(  \underline
{\underline{T}}-\underline{\underline{T}}^{T}\right) \label{to14.1}\\
\underline{\underline{T}}^{sph}  & =\frac{1}{3}\text{tr}\left(  \underline
{\underline{T}}\right)  \underline{\underline{1}}\label{to15}\\
\underline{\underline{T}}^{dev}  & =\underline{\underline{T}}-\underline
{\underline{T}}^{sph},\label{to16}%
\end{align}
and the second-order identity tensor is defined as%
\begin{equation}
\left(  \underline{\underline{1}}\right)  _{ij}=\delta_{ij}.\label{to16.1}%
\end{equation}
where $\delta_{ij}$\ is the Kronecker delta,%
\begin{equation}
\delta_{ij}=\left\{
\begin{array}
[c]{l}%
1\text{ if }i=j\\
0\text{ if }i\neq j
\end{array}
\right\}  .\label{to16.2}%
\end{equation}

\subsubsection*{Differential Operators}

\begin{itemize}
\item gradients (Cartesian coordinate definitions):%
\begin{align}
\left(  \nabla\alpha\right)  _{i}  & =\frac{\partial\alpha}{\partial x_{i}%
}\label{to17}\\
\left(  \nabla\underline{w}\right)  _{ij}  & =\frac{\partial w_{j}}{\partial
x_{i}}\label{to18}%
\end{align}

\item divergences (Cartesian coordinate definitions):%
\begin{align}
\nabla\cdot\underline{w}  & =\frac{\partial w_{i}}{\partial x_{i}}%
\label{to19}\\
\left(  \nabla\cdot\underline{\underline{T}}\right)  _{i}  & =\frac{\partial
T_{ji}}{\partial x_{j}}\label{to20}%
\end{align}

\item Laplacians (Cartesian coordinate definitions):%
\begin{align}
\nabla^{2}\alpha & =\sum_{i=1}^{3}\frac{\partial^{2}\alpha}{\partial x_{i}%
^{2}}\label{to20.5}\\
\left(  \nabla^{2}\underline{w}\right)  _{i}  & =\sum_{j=1}^{3}\frac
{\partial^{2}w_{i}}{\partial x_{j}^{2}}\label{to20.6}%
\end{align}

\item convective derivative:%
\begin{equation}
\frac{D\underline{f}_{\ldots}}{Dt}=\frac{\partial\underline{f}_{\ldots}%
}{\partial t}+\underline{v}\cdot\nabla\underline{f}_{\ldots}\label{to21}%
\end{equation}

\item generalized partial derivative: \ If $\underline{A}_{\ldots}$\ and
$\underline{B}_{\ldots}$\ are tensors of order $M$\ and $N$, respectively,
then%
\begin{equation}
\left(  \frac{\partial\underline{A}_{\ldots}}{\partial\underline{B}_{\ldots}%
}\right)  _{i_{1}\ldots i_{M}j_{1}\ldots j_{N}}=\frac{\partial A_{i_{1}\ldots
i_{M}}}{\partial B_{j_{1}\ldots j_{N}}}.\label{to21.5}%
\end{equation}

\end{itemize}

\subsubsection*{Identities}%

\begin{equation}
\nabla\cdot\left(  \underline{\underline{T}}\cdot\underline{w}\right)
=\underline{w}\cdot\left(  \nabla\cdot\underline{\underline{T}}\right)
+\underline{\underline{T}}\cdot\cdot\nabla\underline{w}=\underline{w}%
\cdot\left(  \nabla\cdot\underline{\underline{T}}\right)  +\underline
{\underline{T}}^{T}\colon\nabla\underline{w}\label{to22}%
\end{equation}%
\begin{equation}
\left(  \underline{u}\,\underline{v}\right)  \cdot\underline{w}=\left(
\underline{v}\cdot\underline{w}\right)  \underline{u}\label{to23}%
\end{equation}%
\begin{equation}
\left(  \underline{u}\,\underline{v}\right)  \cdot\cdot\nabla\underline
{w}=\underline{u}\cdot\nabla\underline{w}\cdot\underline{v}\label{to24}%
\end{equation}%
\begin{equation}
\nabla\cdot\left(  \alpha\underline{A}_{\ldots}\right)  =\alpha\nabla
\cdot\underline{A}_{\ldots}+\left(  \nabla\alpha\right)  \cdot\underline
{A}_{\ldots}\label{to25}%
\end{equation}%
\begin{equation}
\nabla\left(  \underline{u}\cdot\underline{w}\right)  =\left(  \nabla
\underline{u}\right)  \cdot\underline{w}+\left(  \nabla\underline{w}\right)
\cdot\underline{u}\label{to26}%
\end{equation}%
\begin{equation}
\nabla\cdot\left(  \underline{w}\times\underline{\underline{T}}\right)
=\underline{w}\times\left(  \nabla\cdot\underline{\underline{T}}\right)
+C_{1,3}\left[  \left(  \nabla\underline{w}\right)  \times\underline
{\underline{T}}\right] \label{to27}%
\end{equation}%
\begin{equation}
\nabla\cdot\underline{w}=\underline{\underline{1}}\colon\nabla\underline
{w}\label{to28}%
\end{equation}%
\begin{equation}
\nabla\left(  \alpha\beta\right)  =\alpha\nabla\beta+\beta\nabla
\alpha\label{to29}%
\end{equation}%
\begin{equation}
\underline{\underline{T}}^{dev}\colon\underline{\underline{U}}=\underline
{\underline{T}}^{dev}\cdot\cdot\underline{\underline{U}}^{sy,dev}\text{ if
}\underline{\underline{T}}=\underline{\underline{T}}^{T}\label{to30}%
\end{equation}

\begin{equation}
\nabla\cdot\left[  \left(  \nabla\underline{w}\right)  ^{sy,dev}\right]
=\frac{1}{2}\nabla^{2}\underline{w}+\frac{1}{6}\nabla\left(  \nabla
\cdot\underline{w}\right) \label{to31}%
\end{equation}%
\begin{equation}
\nabla\left(  \nabla\cdot\underline{w}\right)  =\nabla^{2}\underline{w}\text{
if }\nabla\times\underline{w}=\underline{0}\label{to32}%
\end{equation}

\subsubsection*{Computations}

\begin{enumerate}
\item Kinetic Energy. \ Taking the partial time derivative of $k$\ as defined
by (\ref{bl7}) and using the product rule, we find%
\[
\frac{\partial k}{\partial t}=\underline{v}\cdot\frac{\partial\left(
\overline{m}\underline{v}\right)  }{\partial t}-\frac{1}{2}\left\Vert
\underline{v}\right\Vert ^{2}\frac{\partial\overline{m}}{\partial t}%
\]
or, substituting (\ref{bl3}) and (\ref{bl1}) into the above,%
\[
\frac{\partial k}{\partial t}=-\underline{v}\cdot\left[  \nabla\cdot\left(
\underline{\underline{P}}+\overline{m}\underline{v}\,\underline{v}\right)
\right]  +\overline{m}\underline{f}_{\overline{M}}\cdot\underline{v}+\frac
{1}{2}\left\Vert \underline{v}\right\Vert ^{2}\nabla\cdot\left(  \overline
{m}\underline{v}\right)  .
\]
One can then use tensor identities (\ref{to22}) and (\ref{to25}) to obtain%
\begin{align*}
\frac{\partial k}{\partial t}  & =-\nabla\cdot\left[  \underline{\underline
{P}}\cdot\underline{v}+\overline{m}\left(  \underline{v}\,\underline
{v}\right)  \cdot\underline{v}\right]  +\underline{\underline{P}}^{T}%
\colon\nabla\underline{v}+\overline{m}\left(  \underline{v}\,\underline
{v}\right)  \cdot\cdot\nabla\underline{v}+\\
& \overline{m}\underline{v}\cdot\underline{f}_{\overline{M}}+\nabla
\cdot\left(  \frac{1}{2}\overline{m}\left\Vert \underline{v}\right\Vert
^{2}\underline{v}\right)  -\frac{1}{2}\overline{m}\underline{v}\cdot
\nabla\left(  \underline{v}\cdot\underline{v}\right)  .
\end{align*}
Finally, if we employ tensor identities (\ref{to23}) and (\ref{to26}), then we
arrive at equation (\ref{bl7.5}).

\item Angular Momentum. \ With the aid of tensor identity (\ref{to27}), the
first term on the right-hand side of (\ref{bl13}) may be written as%
\begin{align}
-\underline{x}\times\left[  \nabla\cdot\left(  \underline{\underline{P}%
}+\overline{m}\underline{v}\,\underline{v}\right)  \right]   & =-\nabla
\cdot\left[  \underline{x}\times\left(  \underline{\underline{P}}+\overline
{m}\underline{v}\,\underline{v}\right)  \right]  +C_{1,3}\left(
\underline{\underline{1}}\times\underline{\underline{P}}\right)  +\nonumber\\
& C_{1,3}\left[  \underline{\underline{1}}\times\left(  \overline{m}%
\underline{v}\,\underline{v}\right)  \right]  ,\label{am1}%
\end{align}
using the fact that%
\[
\nabla\underline{x}=\underline{\underline{1}}%
\]
and the linearity of the contraction operator, $C_{1,3}$. \ Next, by writing
out its components, it is observed that%
\[
C_{1,3}\left(  \underline{\underline{1}}\times\underline{\underline{T}%
}\right)  =\underline{0}%
\]
is satisfied if and only if $\underline{\underline{T}}$\ is symmetric.
\ Therefore, the symmetry of $\overline{m}\underline{v}\,\underline{v}%
$\ causes the third term on the right-hand side of (\ref{am1}) to vanish.
\ Furthermore, the second term also vanishes, implying conservation of
momentum (\ref{bl14}), if and only if $\underline{\underline{P}}$\ is symmetric.
\end{enumerate}

\section{\label{applc}Linear Constitutive Laws}

Here, classical irreversible thermodynamics, as described in de Groot and
Mazur \cite[\textsc{ch. III} and \textsc{IV}]{degroot} and Jou et al.
\cite[\S 1.3]{jou}, is used to derive linear constitutive laws (\ref{ce12}%
)-(\ref{ce14}).

Recall equation (\ref{ce9.3}) for the volumetric entropy production rate and
note that the second term on the right-hand side suggests the following
decomposition for the pressure tensor of a fluid:%
\begin{equation}
\underline{\underline{P}}=P\underline{\underline{1}}+\underline{\underline{P}%
}_{\text{visc}},\label{L1}%
\end{equation}
where $\underline{\underline{P}}_{\text{visc}}$\ represents the viscous part
of the pressure, assumed symmetric, i.e.%
\begin{equation}
\underline{\underline{P}}_{\text{visc}}=\underline{\underline{P}}%
_{\text{visc}}^{T},\label{L2}%
\end{equation}
so that condition (\ref{bl15})\ is satisfied, and $P$, as before, denotes the
thermodynamic pressure. \ Consequently, equation (\ref{ce9.3}) becomes%
\begin{equation}
r_{S}=\underline{q}_{U}\cdot\nabla\left(  \frac{1}{T}\right)  -\frac{1}%
{T}\underline{\underline{P}}_{\text{visc}}\colon\nabla\underline{v}%
-\underline{q}_{M}\cdot\nabla\left(  \frac{\mu}{T}\right)  .\label{L3}%
\end{equation}
One can further decompose the viscous pressure into its spherical and
deviatoric parts via%
\begin{equation}
\underline{\underline{P}}_{\text{visc}}=\underline{\underline{P}}%
_{\text{visc}}^{sph}+\underline{\underline{P}}_{\text{visc}}^{dev}\label{L4}%
\end{equation}
with%
\begin{equation}
\underline{\underline{P}}_{\text{visc}}^{sph}=\widetilde{P}_{\text{visc}%
}\underline{\underline{1}}\text{ where }\widetilde{P}_{\text{visc}}\equiv
\frac{1}{3}\text{tr}\left(  \underline{\underline{P}}_{\text{visc}}\right)
.\label{L5}%
\end{equation}
Substituting (\ref{L4}) and (\ref{L5}) into (\ref{L3}) and using identity
(\ref{to28}) yields%
\begin{align}
r_{S}  & =\underline{q}_{U}\cdot\nabla\left(  \frac{1}{T}\right)
+\widetilde{P}_{\text{visc}}\left(  -\frac{1}{T}\nabla\cdot\underline
{v}\right)  +\nonumber\\
& \underline{\underline{P}}_{\text{visc}}^{dev}\cdot\cdot\left[  -\frac{1}%
{T}\left(  \nabla\underline{v}\right)  ^{sy,dev}\right]  +\underline{q}%
_{M}\cdot\left[  -\nabla\left(  \frac{\mu}{T}\right)  \right]  ,\label{L6}%
\end{align}
where property (\ref{L2}) has been employed in conjunction with tensor
identity (\ref{to30}).

The terms have been grouped as in expression (\ref{L6}) so that $r_{S}$\ has
the form,%
\begin{equation}
r_{S}=\sum_{j}\underline{F}_{\ldots}^{\left(  j\right)  }\left(  \cdot\right)
\underline{A}_{\ldots}^{\left(  j\right)  },\label{L7}%
\end{equation}
where $\underline{F}_{\ldots}^{\left(  j\right)  }$\ and $\underline
{A}_{\ldots}^{\left(  j\right)  }$\ represent fluxes and the affinities,
respectively, and the $\left(  \cdot\right)  $-operator\ is defined in
(\ref{to6.5}). \ One requires the affinities to be objective tensor
quantities\ and for there to exist equilibrium states for which the affinities
and fluxes simultaneously vanish. \ Comparing (\ref{L7}) and (\ref{L6})
suggests the following identifications:%
\begin{equation}%
\begin{tabular}
[c]{|l|l|l|}\hline
$j$ & $\underline{F}_{\ldots}^{\left(  j\right)  }$ (flux) & $\underline
{A}_{\ldots}^{\left(  j\right)  }$ (affinity)\\\hline
$1$ & $\underline{F}^{\left(  1\right)  }=\underline{q}_{U}$ & $\underline
{A}^{\left(  1\right)  }=\nabla\left(  \frac{1}{T}\right)  $\\\hline
$2$ & $F^{\left(  2\right)  }=\widetilde{P}_{\text{visc}}$ & $A^{\left(
2\right)  }=-\frac{1}{T}\nabla\cdot\underline{v}$\\\hline
$3$ & $\underline{\underline{F}}^{\left(  3\right)  }=\underline{\underline
{P}}_{\text{visc}}^{dev}$ & $\underline{\underline{A}}^{\left(  3\right)
}=-\frac{1}{T}\left(  \nabla\underline{v}\right)  ^{sy,dev}$\\\hline
$4$ & $\underline{F}^{\left(  4\right)  }=\underline{q}_{M}$ & $\underline
{A}^{\left(  4\right)  }=-\nabla\left(  \frac{\mu}{T}\right)  $\\\hline
\end{tabular}
.\label{L8}%
\end{equation}
Note that $\underline{A}^{\left(  1\right)  }$, $A^{\left(  2\right)  }$,
$\underline{\underline{A}}^{\left(  3\right)  }$, and $\underline{A}^{\left(
4\right)  }$\ are each indeed objective tensor quantities. \ Next, let us
assume that the fluxes are written as%
\begin{equation}
\underline{F}_{\ldots}^{\left(  j\right)  }=\underline{\widetilde{F}}_{\ldots
}^{\left(  j\right)  }\left(  \underline{A}_{\ldots}^{\left(  1\right)
},\underline{A}_{\ldots}^{\left(  2\right)  },\ldots;\mathcal{T}\right)
,\label{L9}%
\end{equation}
where $\mathcal{T}$\ represents a list of state parameters, each evaluated at
$\left(  \underline{x},t\right)  $, e.g. $\left\{  T,P\right\}  $. \ The
assumption that the affinities and fluxes vanish in a state of equilibrium
implies%
\begin{equation}
\left.  \underline{\widetilde{F}}_{\ldots}^{\left(  j\right)  }\left(
\underline{A}_{\ldots}^{\left(  1\right)  },\underline{A}_{\ldots}^{\left(
2\right)  },\ldots;\mathcal{T}\right)  \right\vert _{\underline{A}_{\ldots
}^{\left(  1\right)  }=\underline{0}_{\ldots},\underline{A}_{\ldots}^{\left(
2\right)  }=\underline{0}_{\ldots},\ldots}=\underline{0}_{\ldots}\label{L10}%
\end{equation}
so that we obtain the following expansions for the fluxes about the
equilibrium state:%
\begin{align}
\underline{F}_{\ldots}^{\left(  j\right)  }  & =\sum_{k}\left.  \frac
{\partial\underline{\widetilde{F}}_{\ldots}^{\left(  j\right)  }}%
{\partial\underline{A}_{\ldots}^{\left(  k\right)  }}\left(  \underline
{A}_{\ldots}^{\left(  1\right)  },\underline{A}_{\ldots}^{\left(  2\right)
},\ldots;\mathcal{T}\right)  \right\vert _{\underline{A}_{\ldots}^{\left(
1\right)  }=\underline{0}_{\ldots},\underline{A}_{\ldots}^{\left(  2\right)
}=\underline{0}_{\ldots},\ldots}\left(  \cdot\right)  \underline{A}_{\ldots
}^{\left(  k\right)  }+\nonumber\\
& (2^{nd}\text{-order terms and higher}).\label{L11}%
\end{align}
Assuming the magnitudes of the affinities are small enough to neglect the
higher-order terms and defining the phenomenological coefficients as%
\begin{equation}
\underline{L}_{\ldots}^{\left(  jk\right)  }\left(  \mathcal{T}\right)
\equiv\frac{\partial\underline{\widetilde{F}}_{\ldots}^{\left(  j\right)  }%
}{\partial\underline{A}_{\ldots}^{\left(  k\right)  }}\left(  \underline
{0}_{\ldots},\underline{0}_{\ldots},\ldots;\mathcal{T}\right)  ,\label{L12}%
\end{equation}
equation (\ref{L11}) becomes the linear constitutive law,%
\begin{equation}
\underline{F}_{\ldots}^{\left(  j\right)  }=\sum_{k}\underline{L}_{\ldots
}^{\left(  jk\right)  }\left(  \mathcal{T}\right)  \left(  \cdot\right)
\underline{A}_{\ldots}^{\left(  k\right)  }.\label{L13}%
\end{equation}
With the use of table (\ref{L8}), the linear constitutive laws for our
single-component viscous fluid are given by%
\begin{align}
\underline{q}_{U}  & =\underline{\underline{L}}^{\left(  11\right)  }%
\cdot\nabla\left(  \frac{1}{T}\right)  +\underline{L}^{\left(  12\right)
}\left(  -\frac{1}{T}\nabla\cdot\underline{v}\right)  +\nonumber\\
& \underline{\underline{\underline{L}}}^{\left(  13\right)  }\cdot\cdot\left[
-\frac{1}{T}\left(  \nabla\underline{v}\right)  ^{sy,dev}\right]
+\underline{\underline{L}}^{\left(  14\right)  }\cdot\left[  -\nabla\left(
\frac{\mu}{T}\right)  \right]  ,\label{L14}%
\end{align}%
\begin{align}
\widetilde{P}_{\text{visc}}  & =\underline{L}^{\left(  21\right)  }\cdot
\nabla\left(  \frac{1}{T}\right)  +L^{\left(  22\right)  }\left(  -\frac{1}%
{T}\nabla\cdot\underline{v}\right)  +\nonumber\\
& \underline{\underline{L}}^{\left(  23\right)  }\cdot\cdot\left[  -\frac
{1}{T}\left(  \nabla\underline{v}\right)  ^{sy,dev}\right]  +\underline
{L}^{\left(  24\right)  }\cdot\left[  -\nabla\left(  \frac{\mu}{T}\right)
\right]  ,\label{L15}%
\end{align}%
\begin{align}
\underline{\underline{P}}_{\text{visc}}^{dev}  & =\underline{\underline
{\underline{L}}}^{\left(  31\right)  }\cdot\nabla\left(  \frac{1}{T}\right)
+\underline{\underline{L}}^{\left(  32\right)  }\left(  -\frac{1}{T}%
\nabla\cdot\underline{v}\right)  +\nonumber\\
& \underline{\underline{\underline{\underline{L}}}}^{\left(  33\right)  }%
\cdot\cdot\left[  -\frac{1}{T}\left(  \nabla\underline{v}\right)
^{sy,dev}\right]  +\underline{\underline{\underline{L}}}^{\left(  34\right)
}\cdot\left[  -\nabla\left(  \frac{\mu}{T}\right)  \right]  ,\label{L16}%
\end{align}
and%
\begin{align}
\underline{q}_{M}  & =\underline{\underline{L}}^{\left(  41\right)  }%
\cdot\nabla\left(  \frac{1}{T}\right)  +\underline{L}^{\left(  42\right)
}\left(  -\frac{1}{T}\nabla\cdot\underline{v}\right)  +\nonumber\\
& \underline{\underline{\underline{L}}}^{\left(  43\right)  }\cdot\cdot\left[
-\frac{1}{T}\left(  \nabla\underline{v}\right)  ^{sy,dev}\right]
+\underline{\underline{L}}^{\left(  44\right)  }\cdot\left[  -\nabla\left(
\frac{\mu}{T}\right)  \right]  .\label{L17}%
\end{align}

As discussed in de Groot and Mazur \cite[\textsc{ch. VI}]{degroot}, in the
absence of external magnetic fields and Coriolis forces, Onsager reciprocity
requires that the phenomenological coefficients appearing in the above satisfy%
\begin{equation}
\underline{L}_{\ldots}^{\left(  jk\right)  }=\left\{
\begin{array}
[c]{c}%
\underline{L}_{\ldots}^{\left(  kj\right)  }\text{ if }\left(  j,k\right)
=\left[
\begin{array}
[c]{c}%
(1,1),(1,4),(2,2),(2,3),\\
(3,2),(3,3),(4,1),\text{ or }(4,4)
\end{array}
\right] \\
-\underline{L}_{\ldots}^{\left(  kj\right)  }\text{ if }\left(  j,k\right)
=\left[
\begin{array}
[c]{c}%
(1,2),(1,3),(2,1),(2,4),\\
(3,1),(3,4),(4,2),\text{ or }(4,3)
\end{array}
\right]
\end{array}
\right.  .\label{L18}%
\end{equation}
Furthermore, as discussed in Segel \cite[\S 2.1]{segel}, isotropic materials
like fluids possess isotropic tensors for each of their phenomenological
coefficients, $\underline{L}_{\ldots}^{\left(  --\right)  }$, and this reduces
equations (\ref{L14})-(\ref{L17}) to%
\begin{equation}
\underline{q}_{U}=\underline{\underline{L}}^{\left(  11\right)  }\cdot
\nabla\left(  \frac{1}{T}\right)  +\underline{\underline{L}}^{\left(
14\right)  }\cdot\left[  -\nabla\left(  \frac{\mu}{T}\right)  \right]
,\label{L24}%
\end{equation}%
\begin{equation}
\widetilde{P}_{\text{visc}}=L^{\left(  22\right)  }\left(  -\frac{1}{T}%
\nabla\cdot\underline{v}\right)  ,\label{L25}%
\end{equation}%
\begin{equation}
\underline{\underline{P}}_{\text{visc}}^{dev}=\underline{\underline
{\underline{\underline{L}}}}^{\left(  33\right)  }\cdot\cdot\left[  -\frac
{1}{T}\left(  \nabla\underline{v}\right)  ^{sy,dev}\right]  ,\label{L26}%
\end{equation}
and%
\begin{equation}
\underline{q}_{M}=\underline{\underline{L}}^{\left(  41\right)  }\cdot
\nabla\left(  \frac{1}{T}\right)  +\underline{\underline{L}}^{\left(
44\right)  }\cdot\left[  -\nabla\left(  \frac{\mu}{T}\right)  \right]
\label{L27}%
\end{equation}
with%
\begin{equation}
L_{ij}^{\left(  11\right)  }=\mathcal{L}^{\left(  11\right)  }\delta
_{ij},\label{L29}%
\end{equation}%
\begin{equation}
L_{ij}^{\left(  14\right)  }=\mathcal{L}^{\left(  14\right)  }\delta
_{ij},\label{L30}%
\end{equation}%
\begin{align}
L_{ij}^{\left(  41\right)  }  & =\mathcal{L}^{\left(  41\right)  }\delta
_{ij}\label{L30.1}\\
& =\mathcal{L}^{\left(  14\right)  }\delta_{ij}\text{ assuming Onsager
reciprocity,}\label{L30.2}%
\end{align}%
\begin{equation}
L_{ij}^{\left(  44\right)  }=\mathcal{L}^{\left(  44\right)  }\delta
_{ij},\label{L31}%
\end{equation}
and%
\begin{equation}
L_{ijkl}^{\left(  33\right)  }=2\mathcal{L}^{\left(  33\right)  }\left[
-\frac{1}{3}\delta_{ij}\delta_{kl}+\frac{1}{2}\left(  \delta_{ik}\delta
_{jl}+\delta_{il}\delta_{jk}\right)  \right] \label{L32}%
\end{equation}
for scalars, $\mathcal{L}^{\left(  11\right)  }$, $\mathcal{L}^{\left(
14\right)  }$, $\mathcal{L}^{\left(  41\right)  }$, $\mathcal{L}^{\left(
44\right)  }$, $\mathcal{L}^{\left(  33\right)  }$, and where $\delta_{ij}%
$\ represents the Kronecker delta.

Substituting (\ref{L29}) and (\ref{L30}) into equation (\ref{L24}), one finds%
\begin{align}
\underline{q}_{U}  & =\mathcal{L}^{\left(  11\right)  }\nabla\left(  \frac
{1}{T}\right)  -\mathcal{L}^{\left(  14\right)  }\nabla\left(  \frac{\mu}%
{T}\right) \nonumber\\
& =\left(  -\frac{\mathcal{L}^{\left(  11\right)  }}{T^{2}}+\frac
{\mu\mathcal{L}^{\left(  14\right)  }}{T^{2}}\right)  \nabla T-\frac
{\mathcal{L}^{\left(  14\right)  }}{T}\nabla\mu,\label{L34}%
\end{align}
whereby defining%
\begin{equation}
k_{U}=\frac{1}{T^{2}}\left(  \mathcal{L}^{\left(  11\right)  }-\mu
\mathcal{L}^{\left(  14\right)  }\right) \label{L35}%
\end{equation}
and%
\begin{equation}
d_{U}=\frac{\mathcal{L}^{\left(  14\right)  }}{\mu T},\label{L35.5}%
\end{equation}
one obtains equation (\ref{ce14}).

Similarly, if one employs (\ref{L30.2}) and (\ref{L31}), then equation
(\ref{L27}) becomes%
\begin{align}
\underline{q}_{M}  & =\mathcal{L}^{\left(  14\right)  }\nabla\left(  \frac
{1}{T}\right)  -\mathcal{L}^{\left(  44\right)  }\nabla\left(  \frac{\mu}%
{T}\right) \nonumber\\
& =\left(  -\frac{\mathcal{L}^{\left(  14\right)  }}{T^{2}}+\frac
{\mu\mathcal{L}^{\left(  44\right)  }}{T^{2}}\right)  \nabla T-\frac
{\mathcal{L}^{\left(  44\right)  }}{T}\nabla\mu\label{L37}%
\end{align}
and, therefore, defining%
\begin{equation}
d_{M}=\frac{\mathcal{L}^{\left(  44\right)  }}{T}\label{L38}%
\end{equation}
and%
\begin{equation}
k_{M}=\frac{\mu}{T^{2}}\left(  \mathcal{L}^{\left(  14\right)  }%
-\mu\mathcal{L}^{\left(  44\right)  }\right)  ,\label{L39}%
\end{equation}
yields equation (\ref{ce12}). \ Note that equations (\ref{L35.5}),
(\ref{L38}), and (\ref{L39}) imply the relation,%
\begin{equation}
k_{M}=\frac{\mu^{2}}{T}\left(  d_{U}-d_{M}\right)  ,\label{L41}%
\end{equation}
which is a consequence of the Onsager reciprocity.

Next, let us substitute (\ref{L32}) into (\ref{L26}) to find%
\begin{align*}
\left(  \underline{\underline{P}}_{\text{visc}}^{dev}\right)  _{ij} &
=-\frac{2\mathcal{L}^{\left(  33\right)  }}{T}\left[  -\frac{1}{3}\delta
_{ij}\delta_{kl}+\frac{1}{2}\left(  \delta_{ik}\delta_{jl}+\delta_{il}%
\delta_{jk}\right)  \right]  \left(  \nabla\underline{v}\right)
_{kl}^{sy,dev}\\
&  =\frac{2\mathcal{L}^{\left(  33\right)  }}{3T}\text{tr}\left[  \left(
\nabla\underline{v}\right)  ^{sy,dev}\right]  \delta_{ij}-\frac{\mathcal{L}%
^{\left(  33\right)  }}{T}\left[  \left(  \nabla\underline{v}\right)
_{ij}^{sy,dev}+\left(  \nabla\underline{v}\right)  _{ji}^{sy,dev}\right]  ,
\end{align*}
which implies%
\begin{equation}
\underline{\underline{P}}_{\text{visc}}^{dev}=-\frac{2\mathcal{L}^{\left(
33\right)  }}{T}\left(  \nabla\underline{v}\right)  ^{sy,dev}\label{L43}%
\end{equation}
since the trace of any deviatoric tensor is zero. \ If one defines%
\begin{equation}
\eta\equiv\frac{\mathcal{L}^{\left(  33\right)  }}{T}\label{L44}%
\end{equation}
and%
\begin{equation}
\zeta\equiv\frac{L^{\left(  22\right)  }}{T},\label{L45}%
\end{equation}
then (\ref{L43}) and (\ref{L25}) become%
\begin{equation}
\underline{\underline{P}}_{\text{visc}}^{dev}=-2\eta\left(  \nabla
\underline{v}\right)  ^{sy,dev}\label{L46}%
\end{equation}
and%
\begin{equation}
\widetilde{P}_{\text{visc}}=-\zeta\nabla\cdot\underline{v}.\label{L47}%
\end{equation}
Finally, using the two equations above, together with (\ref{L5}), (\ref{L4}),
and (\ref{L1}), one arrives at equation (\ref{ce13}) for the total pressure.

\section{\label{appetr}Equilibrium Thermodynamic Relationships}

In addition to the thermodynamic parameters mentioned in \S \ref{secnt}, let
us introduce%
\[%
\begin{tabular}
[c]{ll}%
$h_{M}$ & specific enthalpy\\
$\alpha_{P}$ & coefficient of thermal expansion\\
$\kappa_{T}$ & isothermal compressibility\\
$c_{V}$ & isochoric specific heat per mass
\end{tabular}
.
\]
\ All of these quantities are defined in Callen \cite{callen}.

For thermodynamically stable classical systems, the following inequalities
hold:%
\begin{equation}
m,u,s,P,T,h_{M},\kappa_{T},c_{V}>0\label{e.1}%
\end{equation}
and%
\begin{equation}
\mu<0.\label{e.2}%
\end{equation}
However, there is no sign restriction on $\alpha_{P}$.

Note the following general equilibrium thermodynamic relationships:%

\begin{equation}
h_{M}=\frac{u+P}{m}\label{e10}%
\end{equation}
and%

\begin{align}
dT &  =-\frac{\left(  h_{M}-\frac{T\alpha_{P}}{m\kappa_{T}}\right)  }{mc_{V}%
}dm+\frac{1}{mc_{V}}du\label{e1}\\
dP &  =-\frac{\left(  \alpha_{P}h_{M}-\frac{T\alpha_{P}^{2}}{m\kappa_{T}%
}-c_{V}\right)  }{m\kappa_{T}c_{V}}dm+\frac{\alpha_{P}}{m\kappa_{T}c_{V}%
}du\label{e2}\\
d\mu &  =\left\{
\begin{array}
[c]{c}%
\left[  \frac{1}{m^{2}\kappa_{T}}+\frac{\left(  h_{M}-\frac{T\alpha_{P}%
}{m\kappa_{T}}\right)  \left(  h_{M}-\frac{T\alpha_{P}}{m\kappa_{T}}%
-\mu\right)  }{mTc_{V}}\right]  dm-\\
\frac{\left(  h_{M}-\frac{T\alpha_{P}}{m\kappa_{T}}-\mu\right)  }{mTc_{V}}du
\end{array}
\right\} \label{e3}%
\end{align}
where equations (\ref{e1})-(\ref{e3}) may be derived by using Legendre
transformations, as detailed in Callen \cite[\S 5.3]{callen}.

\section{\label{appmc}Multicomponent Fluid Mixtures}

To model a multicomponent mixture of fluids, one may use the same ideas as
those used to obtain the $\overline{M}\left(  D,\eta\right)  $-formulation for
a single-component fluid. \ Let $M_{i}$\ represent the mass of component
$i$\ in an $n$-component mixture. \ For an $n$-component fluid near
equilibrium and in the hydrodynamic regime, I propose the $\overline{M}%
$-formulation balance laws,%
\begin{align}
\frac{\partial\overline{m}}{\partial t}  & =-\nabla\cdot\left(  \overline
{m}\underline{v}\right) \nonumber\\
\frac{\partial m_{i}}{\partial t}  & =-\nabla\cdot\left(  \underline{q}%
_{M_{i}}+m_{i}\underline{v}\right)  \text{ for }i=1,\ldots,n\label{mf1}\\
\frac{\partial\left(  \overline{m}\underline{v}\right)  }{\partial t}  &
=-\nabla\cdot\left(  \underline{\underline{P}}+\overline{m}\underline
{v}\,\underline{v}\right)  +\overline{m}\underline{f}_{\overline{M}%
}\nonumber\\
\frac{\partial u}{\partial t}  & =-\nabla\cdot\left(  \underline{q}%
_{U}+u\underline{v}\right)  -\underline{\underline{P}}\colon\nabla
\underline{v},\nonumber
\end{align}
with constitutive equations,%
\begin{align}
\underline{q}_{M_{i}}  & =-D\nabla m_{i}\text{ for }i=1,\ldots,n\nonumber\\
\underline{\underline{P}}  & =\left[  P-\left(  \overline{m}D-\frac{4}{3}%
\eta\right)  \nabla\cdot\underline{v}\right]  \underline{\underline{1}}%
-2\eta\left(  \nabla\underline{v}\right)  ^{sy,dev}\label{mf2}\\
\underline{q}_{U}  & =-D\nabla u.\nonumber
\end{align}
Notice that there still appear only two transport coefficients, $D$ and $\eta
$. \ Also, note that since%
\begin{equation}
m=\sum_{i=1}^{n}m_{i}\text{ and }\underline{q}_{M}=\sum_{i=1}^{n}\underline
{q}_{M_{i}},\label{mf3}%
\end{equation}
one of the partial mass equations in the above may be replaced by the total
mass equation,%
\begin{equation}
\frac{\partial m}{\partial t}=-\nabla\cdot\left(  \underline{q}_{M}%
+m\underline{v}\right) \label{mf4}%
\end{equation}
with%
\begin{equation}
\underline{q}_{M}=-D\nabla m.\label{mf5}%
\end{equation}
In this setting, the continuum velocity $\underline{v}$\ is interpreted as a
center of mechanical mass velocity, and in this center of mechanical mass
frame, there is a non-zero total mass flux due to diffusion. \ Also, note that
if the equilibrium thermodynamic pressure, $P$, may be expressed as a function
of $u$\ and $m$\ only--as in the case of a homogeneous mixture--the
$n-1$\ partial mass density equations may potentially be uncoupled from the
rest of the system, leaving the $\overline{M}\left(  D,\eta\right)
$-formulation derived for a single-component fluid. \ This means that
homogeneous mixtures such as air may be treated in the same way as a
single-component fluid.

\end{document}